\documentclass[conference]{IEEEtran}
\IEEEoverridecommandlockouts
\usepackage{amsmath, amssymb, amsfonts, graphicx, algorithm, algorithmic, enumerate, color}
\usepackage{cite}
\usepackage{textcomp}
\usepackage{mathtools}
\usepackage{longtable}
\usepackage[outdir=./]{epstopdf}
\usepackage{graphicx}

\usepackage{tikz}
\newcommand{\comment}[1]{}

\def\BibTeX{{\rm B\kern-.05em{\sc i\kern-.025em b}\kern-.08em
    T\kern-.1667em\lower.7ex\hbox{E}\kern-.125emX}}

\begin{document}

\title{Interference and noise cancellation for joint communication radar (JCR) system based on contextual information\\

\thanks{This work was supported in part by the UK EPSRC under grants EP/P009670/1 and EP/T021063/1, and Petroleum Technology Development Fund (Grant number: PTDF/ED/PHD/NCO/1352/18).}}

\author{\IEEEauthorblockN{Christantus O. Nnamani,}~\IEEEmembership{Member,~IEEE}
\IEEEauthorblockA{\textit{School of Aerospace, Transport and Manufacturing} \\
\textit{Cranfield University}\\
Bedford, UK \\
christantus.nnamani@cranfield.ac.uk}
\and
\IEEEauthorblockN{Mathini Sellathurai,}~\IEEEmembership{Senior Member,~IEEE}
\IEEEauthorblockA{\textit{Engineering and Physical Sciences} \\
\textit{Heriot-Watt University}\\
Edinburgh, UK \\
m.sellathurai@hw.ac.uk}
}

\maketitle

\begin{abstract} \label{abs}
This paper examines the separation of wireless communication and radar signals, thereby guaranteeing cohabitation and acting as a panacea to spectrum sensing. First, considering that the channel impulse response was known by the receivers (communication and radar), we showed that the optimizing beamforming weights mitigate the interference caused by signals and improve the physical layer security (PLS) of the system. Furthermore, when the channel responses were unknown, we designed an interference filter as a low-complex noise and interference cancellation autoencoder. By mitigating the interference on the legitimate users, the PLS was guaranteed. Results showed that even for a low signal-to-noise ratio, the autoencoder produces low root-mean-square error (RMSE) values.
\end{abstract}

\section{Introduction}\label{intro}
Due to advances in vehicular infrastructure and the need for driverless vehicles, studies into the feasibility of the cohabitation of various sensors operating at diverse spectrum bands began to emerge \cite{Martone_2021, Jarmo_2009, Nasser_2021}. This was further exacerbated by the congestion of the below 6GHz spectrum band mainly used for low earth spectrum applications. A prominent sandwich of application in most discourse is radar and wireless communication (herein referred to as communication) applications with critical reviews presented in \cite{Mazahir_2021, Martone_2021}. Both spectrum, in principle can collaborate via cohabitation, co-design and cooperation \cite{Mazahir_2021}. However, the collaboration is marred by several design challenges such as interference management, varying power requirements, integration and security. A typical paradigm presents that radar signals require higher transmit power than the conventional wireless communication signals. However, the reflected radar signal used in target assessment is considered low powered making it highly susceptible to interference from communication signals. Such exemplar describes the need for cohabitation for both signals to optimize the usage of the available spectrum.

Cohabitation of radar and communication signals are broadly discussed under the dual-function radar communication (DFRC) and joint communication and radar (JCR) models using multiple input multiple output (MIMO) systems. While the latter presents complementary roles for both signals, the former describes a waveform that inseparably represents both signals. Overviews of the coexistence of communication and radar systems were presented in \cite[part 1]{Liu_2020}, \cite{Zhang_2021, Thoma_2021}. 

A trade-off analysis for the conflicting requirements of power and signal space for a JCR half-duplex system was addressed in \cite{Li_2021}. In \cite[part 2]{Liu_2020}, a scheme that estimates the communication channel while conducting radar target detection was proposed. The scheme use hybrid-analog-digital (HAD) beamformer to transmit pilot signals for channel estimation and target searching. Similarly, an interweave full-duplex co-existence scheme was presented in \cite{Sodagari_2012}, where the radar signal was projected onto the null space of the channel matrix between the radar and communication signals. Soft physical layer security (PLS) guarantees cannot be obtained for the JCR systems due to the exposure of the communication signals to the radar target(s) and receivers. 
PLS of the cohabitation of radar and communication systems considers that the radar targets and/or receivers may likely be unintended receivers of communication signals eavesdropping on the transmission.

In the DFRC, embedded communication signals was performed on the beamforming weights or on the orthogonal waveform or vice versa \cite{Hassanien_2018}. Emphasizing on maintaining power levels and maximizing signal-to-interference noise ratio (SINR), the beam pattern obtained from the co-variance matrix of the radar signal was used to obtain the transmit beamforming through zero-forcing precoding \cite{Liu_2018}. Therefore, beamforming designs for full and/or half-duplex transmit and receive communication mitigates the interference between radar and communications signals at the expense of the PLS of the communication signals. However, to ameliorate the PLS concerns, the communication signal and some artificial noise (AN) were embedded onto the beamforming weights of the radar transmission \cite{Su_2019}. Furthermore, the DFRC system was implemented by using the main lobe for radar and the sidelobes for communication transmissions \cite{Hassanien_2016}. The communication signals were embedded in the signal waveforms determined by 2 different beamforming weights (representing 0 and 1). Although attempts were made by \cite{Su_2019} and \cite{Hassanien_2016} to incorporate PLS in DFRC system, they were transmission-centred, based on statistical knowledge of the channel impulse and required handshake between the communication transmitter and genuine receivers.

Nevertheless, if real-time channel information and/or noise impact are unavailable PLS and interference management challenges become exacerbated. This is further worsened when the establishment of communication handshake is impossible. Considering autonomous multi-application domain, it is apparent that the receiver systems become equipped with interference cancellation abilities to maximize the quality of received signal and improve on the PLS of the communication. 
\begin{enumerate}
    \item We first evaluate the performance of the communication and radar cohabiting system when its design allow for the cancellation of the interfering signals to both receivers. This was achieved by assuming the channel information are available and nulling the interfering transmissions with beamforming weights.
    \item Furthermore, relaxing the assumption on the channel information (i.e. by considered that the channel information are unknown), this paper propose an interference mitigation scheme implemented with autoencoders. We focus on separating transmitted communication and radar signals at the receiver of the communication and the radar system using novel noise and interference cancellation filter. Our novel approach entails the use of autoencoders at the receivers to filter out the interfering and noise signals. We note that the proposed method curtails the requirement for spectrum sensing since the impact of the interfering signal is minimized in terms of SINR. By limiting the spurious interfering communication signal impinging on the legitimate receivers while confusing the eavesdroppers, PLS performance was improved. Similarly, by reducing the radar interference, the communication transmission rate was also improved. 
\end{enumerate}

In practice, an application of the proposed noise and interference cancellation filter aids autonomous vehicular radar impact on wireless communication infrastructure. Although it minimizes the impact of the DFRC and/or JCR constraints, we focus on the cohabiting of JCR system without loss of generalization. We emphasize that the overall objective entails the cancellation of interference at legitimate receivers based on previously acquired contextual information of the system's reaction to cohabitation. Such contextual information as applicable to radar tracking system using neural networks \cite{Antonio_2013} were implemented. Specifically, the contextual information is obtained from an \textit{a priori} knowledge of observations of the JCR system with pilot/test sample signals. 


\textit{Notations:}
The structure of the notations employed in this paper elucidate that $\{\cdot\}^{*}$, $\{\cdot\}^{\rm T}$ and $\{\cdot\}^{\rm H}$ represent the conjugate, transpose and Hermitian of vectors/matrices, respectively. Low case letters are scalars, bold-faced low case letters are vectors while bold-faced upper case letters are matrices. Furthermore, rank$({\bf X})$ and Tr$({\bf X})$ are the rank and trace of matrix ${\bf X}$ respectively.

\section{System Model of a Communication/Radar Cohabiting Scenario}\label{sec_sys_model5}
\begin{figure}[!ht]
\centering
\includegraphics[width=1.0\linewidth]{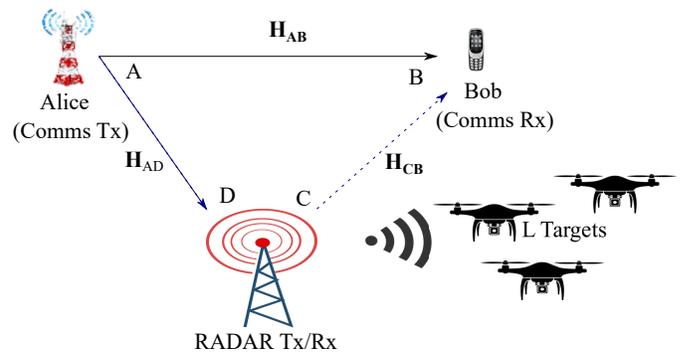}
\caption{MIMO Communication and radar cohabitation system}
\label{case_2_paper5}
\end{figure}

Consider a MIMO communication system with $N_{\rm A}$ transmit and $N_{\rm B}$ receive antennas operating at the radar spectrum. Let the communication system cohabit with MIMO radar systems with $N_{\rm D}$ transmit and $N_{\rm C}$ receive antennas. 
For computational simplicity, we assume that the transmit and receive radar systems are located at the same place such that they share the same far-field observations. However the assumption does not necessarily apply to the communication system, thereby supporting their joint action. The $N_{\rm D}\times 1$ and $N_{\rm A}\times 1$ passband signals of the radar and communication transmit antennas are presented in \eqref{baseband_RADAR5} and \eqref{baseband_comm5} respectively.
\begin{subequations}
\begin{align}
    {\bf s}_{m}(t)={\bf w}^*_m\psi_m(t), ~\forall{~m=\{1,\dots,M\}} \label{baseband_RADAR5}\\
    {\bf s}_{{\rm AB}}(t)={\bf w}_{{\rm AB}}^*\varphi(t), ~\forall{~t=\{1,\dots,T\}.} \label{baseband_comm5}
\end{align}
\end{subequations}
where ${\bf w}_k$ represent the $N_{\rm D} \times 1$ transmit beamforming vector of the $m$th orthogonal waveform ($\psi_m(t)$). ${\bf w}_{{\rm AB}}$ the $N_{\rm A}\times 1$ is the transmit weight vector of the communication systems. $M$ is the total number of orthogonal waveforms radiated by the radar system (for simplicity, $M=N_{\rm D}$ implying that the radar transmission are orthogonal for each antenna) and $T$ is the total number of transmit snapshots. While $\varphi(t)$ is the baseband communication signal waveform. If we assume frequency hopping communication transmission with quadrature phase shift keying (PSK) modulation, then $\varphi(t)$ is presented in \cite[eq. 9]{Hassanien_2018}. 
In the rest of this paper, the subscripts ${\rm A,B,C,D, E}$ denotes an index of the communication transmitter, communication legitimate receiver (Bob), radar transmitter, radar receiver and communication illegitimate receiver (Eve) respectively. We note that in the context of JCR, the communication and radar transmit systems are usually co-located, sharing the same physical resources. However, spatial separation has been introduced in fig.~\ref{case_2_paper5} for clarity.

To understand the operational requirements of the entire system model, we discuss 2 distinct role of the radar system in relation to the communication system. The distinct roles care considered when radar target is absent and when it is present.

\subsection{Case 1} \label{sec_case15}
In this section, we model the received signals of the communications and radar receiver systems under JCR when they are no radar targets. The received signal at the communication and the radar receivers are given as \eqref{comm_rx15} and \eqref{RADAR_rx15} respectively. 
\begin{subequations}
\begin{align}
    {\bf y}_i(t)=&\underbrace{{\bf H}_{{\rm A}i}{\bf s}_{{\rm AB}}(t)}_{\rm Comms.~transmitted} 
    +\underbrace{{\bf H}_{{\rm C}i}{\bf s}_{m}(t)}_{\rm radar~interfered}+{\bf n}_i, \label{comm_rx15} \\
    {\bf y}_{\rm D}(t)=&\underbrace{{\bf H}_{\rm AD}{\bf s}_{\rm AB}(t)}_{\rm Comms.~transmitted}+{\bf n}_{\rm D},\label{RADAR_rx15}
\end{align}
\end{subequations}
$~\forall{~i \in \{{\rm B,E}\}}$, and where ${\bf H}_{{\rm A}i}={\bf b}(\theta_{i{\rm A}})\alpha_{{\rm A}i}{\bf a}^{\rm T}(\theta_{{\rm AB}})$, ${\bf H}_{\rm AD}={\bf d}(\theta_{\rm DA})\alpha_{{\rm AD}}{\bf a}^{\rm T}(\theta_{\rm AB})$, ${\bf H}_{{\rm C}i}={\bf b}(\theta_{i{\rm C}})\alpha_{{\rm C}i}{\bf c}^{\rm T}(\theta)$. $\alpha_{jk}=\rho_0\zeta\|\boldsymbol{\Omega}_j-\boldsymbol{\Omega}_k\|^{-2}, ~\forall{~\{j~ \& ~k\in\{{\rm A,B,C,D,E}\}\}}$ are random channel coefficients characterizing the propagation from path $j$ to $k$. $\rho_0$ represents the channel power gain at reference distance $d_0=1$ m and $\zeta$ is an exponential random variable with unit mean similar to \cite{uav_cooperative_jamming, uav_secured_com_JTTP, obinna_2022}. Parameters
${\bf a}(\theta_{\rm AB})$ and ${\bf c}(\theta)$ are the transmit communication and radar steering vectors respectively. While ${\bf b}(\theta_{i{\rm A}})$, ${\bf b}(\theta_{i{\rm C}})$ and ${\bf d}(\theta_{\rm DA})$ are the receive steering vectors for the communication and radar systems. Note that the communication receivers include the legitimate receiver (Bob) and the eavesdropper (Eve) as required. Assume that the antenna geometry on the communication and radar systems follow a uniform linear array (ULA) configuration, then the steering vectors was generated with
$${\bf x}(\theta_{jk})=[1, e^{j\frac{2\pi}{\lambda}d_x\sin(\theta_{jk})}, \dots,e^{j\frac{2\pi}{\lambda}(N_{i}-1)d_x\sin(\theta_{jk})}]^{\rm T},$$
where $d_x$ is the distance between antenna elements. Furthermore, ${\bf n}_{\rm B}\sim \mathcal{CN}(0,\sigma_{\rm B}^2{\bf 1}_{N_{\rm B}})$ and ${\bf n}_{\rm D} \sim \mathcal{CN}(0,\sigma_{\rm D}^2{\bf 1}_{N_{\rm C}})$ are additive white Gaussian noise with variance $\sigma_{\rm B}^2$ and $\sigma_{\rm D}^2$. The spatial direction of the radar system is focused towards a predefined sector such that $\theta =[\Theta_{\rm min},\Theta_{\rm max}]$ where $\Theta_{\rm min}$ and $\Theta_{\rm max}$ represents the lower and upper contours of the sector. We note that $\theta_{jk}$ is the azimuth spatial direction of transmission from the $j$ to the $k$ or reception at $j$ from $k$. 

\subsection{Case 2} \label{sec_case25}
Consider fig.~\ref{case_2_paper5} where hypothetical $L$ radar targets reflects radar signals. The received signal equations at the communication and radar receiver are given in \eqref{comm_rx25} and \eqref{RADAR_rx25} respectively.

\begin{subequations}
\begin{align}
    {\bf y}_i(t)=&{\bf H}_{{\rm A}i}{\bf s}_{{\rm AB}}(t)
    +{\bf H}_{{\rm C}i}{\bf s}_{m}(t) 
    +\underbrace{\sum_{l=1}^L{\bf H}_{li}{\bf s}_{m}(t)}_{\rm target~reflected}+{\bf n}_i,  \label{comm_rx25} \\
    {\bf y}_{\rm D}(t)=&{\bf H}_{\rm AD}{\bf s}_{{\rm AB}}(t) 
    +\underbrace{\sum_{l=1}^L{\bf H}_{l{\rm D}}{\bf s}_{m}(t)}_{\rm target~reflected}+{\bf n}_{\rm D}, \label{RADAR_rx25}
\end{align}
\end{subequations}
$~\forall{~i \in \{{\rm B,E}\}}$, and where ${\bf H}_{li}={\bf b}(\theta_{il})\alpha_{li}\beta_l\alpha_{lr}{\bf c}^{\rm T}(\theta)$, ${\bf H}_{l{\rm D}}={\bf d}(\theta_{{\rm D}l})\alpha_{l{\rm D}}\beta_l\alpha_{lr}{\bf c}^{\rm T}(\theta)$.
$\beta_l$ obeys Swerling II model and represents the reflection coefficient of the $l$th target. The Swerling II model imply that the reflectivity of the target will change for different pulse, but it will be constant within the duration of a single pulse duration.

\textit{Section summary:} In the cases described in \ref{sec_case15} and \ref{sec_case25}, it is underlined that some form of cross-interference exist for the communication and radar systems especially when their channels are correlated. For the communication signal containing relevant data, its interference on the radar receiver makes it susceptible to eavesdropping. By mitigating the interference, the loss of data in the physical layer domain is reduced. 

\section{Interference Mitigation}\label{sec_intef_mit5}
In this section, we examine the methods to mitigate the interference caused by the communication transmission on the radar reception and vice versa. The interference mitigation approaches are discussed under two distinct generic scenarios, namely cooperative and uncooperative systems.
\subsection{Cooperative Systems}\label{cooperative5}
The radar and communication systems are cooperative when the channel impulse responses between the transmitters and receivers are known by both systems. This entails that channel estimation had been carried out and updated in both systems. Using this contextual information of the \textit{a priori} channel information, the interferring signals are reduced in the transmitter design. 
It is easy to see that by independently optimizing the transmit beamforming weights of both the communication and radar transmissions, interference cancellation is obtained in the case discussed in section \ref{sec_case15}. However, the design of the beamforming weights do not cancel the interference caused by radar target reflection as in the case discussed in section \ref{sec_case25}. To buttress the theoretical formulation, we formulate the optimization problems in \eqref{co_com_int5} and \eqref{co_rad_int5}. These equations were independently solved at the communication and radar transmitters respectively.

\begin{subequations}\label{co_com_int5}
\begin{align}
    \max_{{\bf w}_{{\rm AB}}} ~&\log_2\big(1+{\gamma}_{\rm B}\big), \\
    {\rm s.t.} ~&{\rm Tr}(|{\bf H}_{\rm AD}{\bf s}_{{\rm AB}}(t)|^2)=0, \label{null_com5}\\
    ~& {\bf w}_{\rm AB}^{\rm H}{\bf w}_{\rm AB}=1, \label{com_power5}
\end{align}
\end{subequations}
where ${\gamma}_{\rm B}$ represents the SINR at the legitimate receiver (Bob) are its expression given in \eqref{sinr_b5}.
\begin{align}
{\gamma}_{\rm B} & =\frac{{\rm Tr}(|{\bf H}_{\rm AB}{\bf s}_{{\rm AB}}(t)|^2)}{\sum_{l=1}^L{\rm Tr}(|{\bf H}_{l{\rm B}}{\bf s}_{m}(t)|^2)+{\rm Tr}(|{\bf H}_{{\rm CB}}{\bf s}_{m}(t)|^2)+\sigma^2_{\rm B}}. \label{sinr_b5}
\end{align}

\begin{subequations}\label{co_rad_int5}
\begin{align}
    \max_{{\bf w}_k} ~&\log_2\Bigg(1+\frac{\sum_{l=1}^L{\rm Tr}(|{\bf H}_{l{\rm D}}{\bf s}_{m}(t)|^2)}{{\rm Tr}(|{\bf H}_{\rm AD}{\bf s}_{{\rm AB}}(t)|^2)+\sigma^2_{\rm D}}\Bigg), \\
    {\rm s.t.} 
   ~& \sum_{l=1}^L{\rm Tr}(|{\bf H}_{l{\rm B}}{\bf s}_{m}(t)|^2)+{\rm Tr}(|{\bf H}_{\rm CB}{\bf s}_{m}(t)|^2)=0, \label{null_rad5}\\
    ~& {\bf w}_{k}^{\rm H}{\bf w}_{k}=1 \label{rad_power5}.
\end{align}
\end{subequations}
Recalling that ${\bf s}_{m}(t)$ and ${\bf s}_{{\rm AB}}(t)$ were previously defined as a function of the beamforming weights in \eqref{baseband_RADAR5} and \eqref{baseband_comm5} respectively, the solution to \eqref{co_com_int5} and \eqref{co_rad_int5} are easily obtained using cvx \cite{cvx2}. 
We note that \eqref{com_power5} and \eqref{rad_power5} describes the total power transmitted by the communication and radar transmitter respectively. These powers can be scaled to desired level in practice. For $L=0$, no target reflects the radar signal and the solution is produced for section \ref{sec_case15}.

\subsection{PLS Analysis of the Communication/Radar Cohabiting Scenario} \label{pls_radar5}
\begin{figure}[H]
\centering
\includegraphics[width=1.0\linewidth]{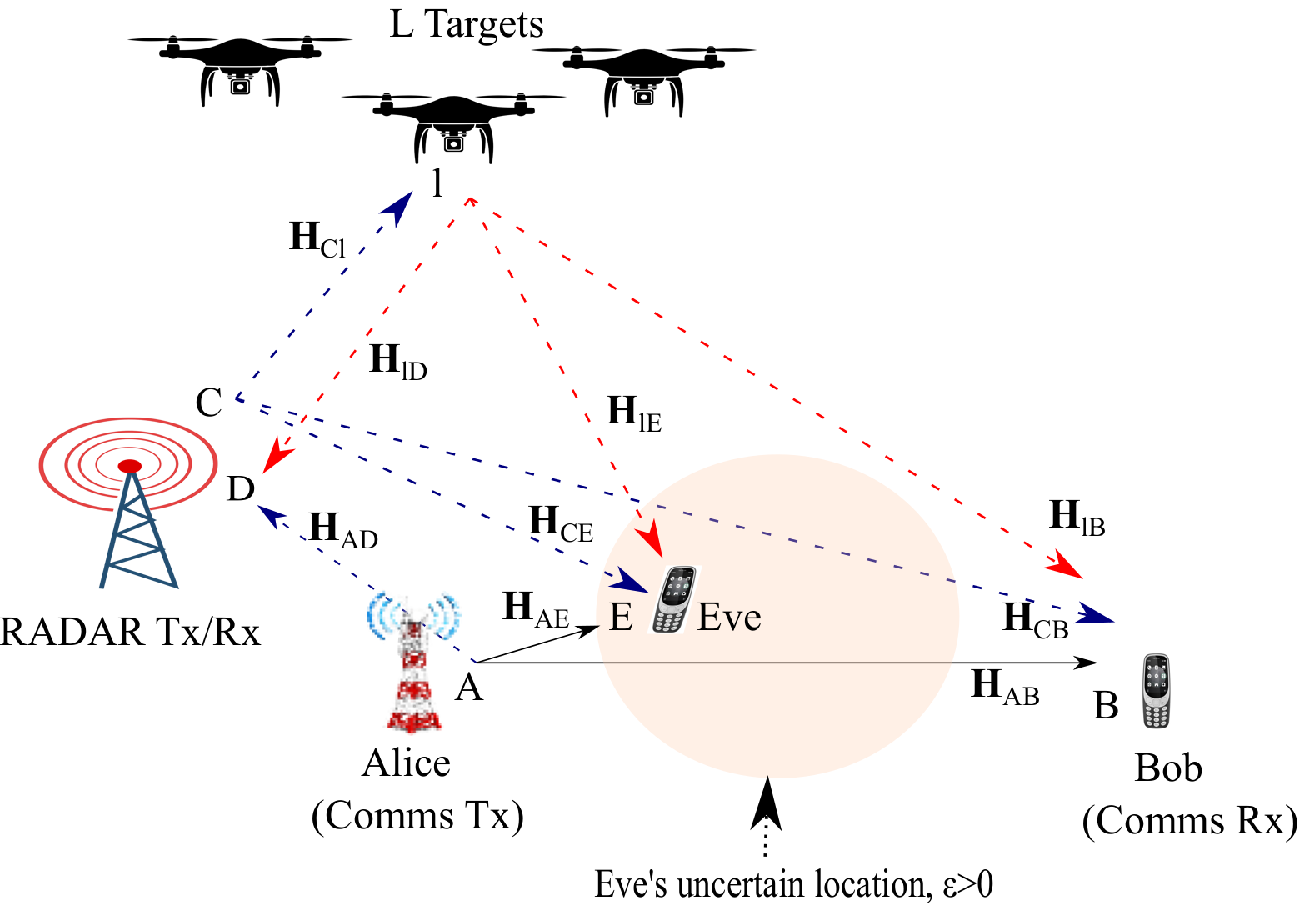}
\caption{MIMO Communication and radar cohabitation system with interaction from a passive eavesdropper}
\label{case_25}
\end{figure}

Consider that an eavesdropper lurks within the radio vicinity of the wireless communication signal, thereby receiving the legitimate communication transmission and interference generated by the cohabiting transmission (radar). In this section, the characterization of the PLS with communication sensing was performed. The generic cohabitation figure presented in fig.~\ref{case_2_paper5} was expanded in fig.~\ref{case_25} with the depiction of the passive eavesdropper location to allow for PLS analysis. 

Since the eavesdropper is passive, its exact location or signal signature is unknown. However, for simplicity, we assume that it is located within a circular region that spans the coverage area of the transmitters.  Therefore, following the derivations in \cite{Nnamani_2021}, the exact location of Eve ($\boldsymbol{\Omega}_{\rm E}$) was defined as a point on a circular uncertain region with uncertainty given in \eqref{Eve_loc5}.
\begin{subequations}\label{Eve_loc5}
\begin{align} 
& \boldsymbol{\Omega}_{\rm E}=\hat{\boldsymbol{\Omega}}_{\rm E}\pm \Delta \boldsymbol{\Omega}_{\rm E},  \label{Eve_loca5}\\
& \|\pm \Delta \boldsymbol{\Omega}_{\rm E}\|=\|\boldsymbol{\Omega}_{\rm E}-\hat{\boldsymbol{\Omega}}_{\rm E}\|\leq \varepsilon, ~{\rm for}~ \varepsilon \geq 0,  \\
& \| \Delta \boldsymbol{\Omega}_{\rm E}\| \leq \varepsilon,
\end{align}
\end{subequations}
holds true, where $\hat{\boldsymbol{\Omega}}_{\rm E}$, $\Delta \boldsymbol{\Omega}_{\rm E}$ and $\varepsilon$ define the estimated location of Eve, the error of the estimation and the radius of error, respectively. 
The estimated location of Eve influences its channel coefficient previously defined as $\alpha_{j{\rm E}}=\rho_0\zeta\|\boldsymbol{\Omega}_j-\boldsymbol{\Omega}_{\rm E}\|^{-2}$, where $j\in\{{\rm A,C,l}\}$. Using triangular inequality and substituting \eqref{Eve_loc5}, we have that
\begin{align} \label{dist_cons5}
&\|\boldsymbol{\Omega}_j-\boldsymbol{\Omega}_{\rm E}\|=\|\boldsymbol{\Omega}_j-(\hat{\boldsymbol{\Omega}}_{\rm E}\pm\Delta \boldsymbol{\Omega}_{\rm E})\| \leq \|\boldsymbol{\Omega}_j-\hat{\boldsymbol{\Omega}}_{\rm E}\|+\varepsilon.
\end{align}
The right hand side is a upper bound to euclidean distance between the transmitters (communication, radar) and the center of the circular uncertain region. By substituting the approximation of the upper bound of the  location of Eve, we obtain that $\hat{\alpha}_{j{\rm E}}=\rho_0\zeta\big(\|\boldsymbol{\Omega}_j-\boldsymbol{\hat{\Omega}}_{\rm E}\|+\varepsilon\big)^{-2}$. Hence the SINR of Eve was presented in \eqref{sinr_e5} respectively.
\begin{align}
\hat{\gamma}_{\rm E} & = \frac{{\rm Tr}(|{\bf H}_{\rm AE}{\bf s}_{{\rm AB}}(t)|^2)}{\sum_{l=1}^L{\rm Tr}(|{\bf H}_{l{\rm E}}{\bf s}_{m}(t)|^2)+{\rm Tr}(|{\bf H}_{\rm CE}{\bf s}_{m}(t)|^2)+\sigma^2_{\rm E}}, \label{sinr_e5}
\end{align}
where ${\bf H}_{\rm AE}={\bf b}(\theta_{\rm EA})\hat{\alpha}_{{\rm AE}}{\bf a}^{\rm T}(\theta_{\rm AB})$, ${\bf H}_{l{\rm E}}={\bf b}(\theta_{{\rm E}l})\alpha_{l{\rm E}}\beta_l\hat{\alpha}_{lr}{\bf c}^{\rm T}(\theta)$. Therefore, the average secrecy rate is the difference in the information rate of Bob and Eve is given in \eqref{rs5} \cite{on_the_secrecy,uav_cooperative_jamming}.
\begin{equation} \label{rs_main5}
R_s = \big[\log_2(1+\gamma_{\rm B})-\log_2(1+\hat{\gamma}_{\rm E})\big]^+,
\end{equation}
where $[x]^+={\rm max}\{0,x\}$ ensures that the information rate received by Eve is not greater than that received by Bob, therefore guaranteeing positive average secrecy rates \cite{on_the_secrecy}. However, for ease of computation, this constraint on the average secrecy rate is ignored. This was justified with the analysis that the system provides for degrading of the eavesdropper's channel leading to positive secrecy rates. 

%
\begin{subequations}\label{co_com_int_15}
\begin{align}
    \max_{{\bf w}_{{\rm AB}},{\bf w}_{k}} ~& R_s, \label{rs5} \\
    {\rm s.t.} ~&\log_2\Bigg(1+\frac{\sum_{l=1}^L{\rm Tr}(|{\bf H}_{l{\rm D}}{\bf s}_{m}(t)|^2)}{\sigma^2_{\rm D}}\Bigg)\geq r_{\rm th}, \label{radar_rate_threshold_15} \\
    ~&{\rm Tr}(|{\bf H}_{\rm AD}{\bf s}_{{\rm AB}}(t)|^2)=0, \label{null_com_15}\\
    ~& \sum_{l=1}^L{\rm Tr}(|{\bf H}_{l{\rm B}}{\bf s}_{m}(t)|^2)+{\rm Tr}(|{\bf H}_{{\rm CB}}{\bf s}_{m}(t)|^2)=0, \label{com_int_canc_15} \\
    ~& {\bf w}_{\rm AB}^{\rm H}{\bf w}_{\rm AB}=1, \label{com_power_15}\\
    ~& {\bf w}_{k}^{\rm H}{\bf w}_{k}=1 \label{rad_power_15}.
\end{align}
\end{subequations}
The parameter $r_{\rm th}$ is the maximum radar rate required to reformulate the reflected signal from the radar targets. Equation \eqref{radar_rate_threshold_15} provides the lower bound to the rate received by the radar receiver to reconstruct the reflected signal. Equation \eqref{com_int_canc_15} used the known channel information to cancel the interference at the legitimate receiver while \eqref{null_com_15} removes the interference of the communication signal at the radar receiver. By substituting for ${\bf s}_{\rm AB}$ and ${\bf s}_{\rm AB}$ with \eqref{baseband_comm5} and \eqref{baseband_RADAR5} respectively and expanding the objective function, \eqref{co_com_int_15} is rewritten as \eqref{co_com_int_1b5}.
\begin{subequations} \label{co_com_int_1b5} 
\begin{align} 
    \max_{{\bf W}_{{\rm AB}},{\bf W}_{k}} ~&\log_2\big(1+\gamma_{\rm B}\big)
    -\log_2\big(1+\hat{\gamma}_{\rm E}\big), \label{rsb5} \\
    {\rm s.t.} ~&\log_2\Bigg(1+\frac{\sum_{l=1}^L{\rm Tr}({\bf H}_{l{\rm D}}{\bf W}_{k}{\bf H}_{l{\rm D}}^{\rm H})}{\sigma^2_{\rm D}}\Bigg)\geq r_{\rm th}, \label{radar_rate_threshold_1b5} \\
    ~&{\rm Tr}({\bf H}_{\rm AD}{\bf W}_{{\rm AB}}{\bf H}_{\rm AD}^{\rm H})=0, \label{null_com_1b5}\\
    ~& \sum_{l=1}^L{\rm Tr}({\bf H}_{l{\rm B}}{\bf W}_{k}{\bf H}_{l{\rm B}}^{\rm H})+{\rm Tr}({\bf H}_{{\rm CB}}{\bf W}_{k}{\bf H}_{{\rm CB}}^{\rm H})=0, \label{com_int_canc_1b5} \\
    ~& {\rm Tr}({\bf W}_{\rm AB})=1, \label{com_power_1b5}\\
    ~& {\rm Tr}({\bf W}_{k})=1, \label{rad_power_1b5}\\
    ~& {\rm rank}({\bf W}_{\rm AB})=1, \label{com_rank_1b5}\\
    ~& {\rm rank}({\bf W}_{k})=1. \label{rad_rank_1b5}
\end{align}
\end{subequations}
Equations \eqref{com_rank_1b5} and \eqref{rad_rank_1b5} were consequences of ${\bf W}_{\rm AB}={\bf w}_{\rm AB}{\bf w}_{\rm AB}^{\rm H}$, and ${\bf W}_{k}={\bf w}_{k}{\bf w}_{k}^{\rm H}$. The SINR equation given in \eqref{sinr_b5}, with the interference nulling performed in \eqref{com_int_canc_1b5}, and the SINR of \eqref{sinr_e5} were expanded to
$$\gamma_{\rm B}=\frac{{\rm Tr}({\bf H}_{\rm AB}{\bf W}_{{\rm AB}}{\bf H}_{\rm AB}^{\rm H})}{\sigma^2_{\rm B}}.$$
$$\hat{\gamma}_{\rm E}=\frac{{\rm Tr}({\bf H}_{\rm AE}{\bf W}_{{\rm AB}}{\bf H}_{\rm AE}^{\rm H})}{\sum_{l=1}^L{\rm Tr}({\bf H}_{l{\rm E}}{\bf W}_{k}{\bf H}_{l{\rm E}}^{\rm H})
+{\rm Tr}({\bf H}_{\rm CE}{\bf W}_{k}{\bf H}_{\rm CE}^{\rm H})+\sigma^2_{\rm E}}.$$

We note that \eqref{co_com_int_1b5} is non-convex due to the non-convexity of the objective function. However, it was solved by applying successive convex approximation (SCA). The SCA allows the problem to be broken into sub-optimal problems and an iterative algorithms developed to minimize the error of the objective function given in \eqref{rsb5} at each iteration step. The sub-problems and solutions arising from \eqref{co_com_int_1b5} were presented as \eqref{co_com_int_115} and \eqref{co_com_int_125} and the iterative algorithm was summarized in Algorithm~\ref{algo15}.

First, we present the sub-problem from \eqref{co_com_int_1b5} that solves for the beamforming weights parameter arising from the wireless communication transmission in \eqref{co_com_int_115}.
\begin{subequations}\label{co_com_int_115}
\begin{align}
    \max_{{\bf W}_{{\rm AB}}} ~&\log_2\big(1+{\bar{k}_1{\rm Tr}({\bf H}_{\rm AB}{\bf W}_{{\rm AB}}{\bf H}_{\rm AB}^{\rm H})}\big) \nonumber\\
    ~&-\log_2\big(1+{\bar{k}_2{\rm Tr}({\bf H}_{\rm AE}{\bf W}_{{\rm AB}}{\bf H}_{\rm AE}^{\rm H})}\big), \label{rs15} \\
    {\rm s.t.} ~&{\rm Tr}({\bf H}_{\rm AD}{\bf W}_{{\rm AB}}{\bf H}_{\rm AD}^{\rm H})=0, \label{null_com_115}\\
    ~& {\rm Tr}({\bf W}_{\rm AB})=1, \label{com_power_115}\\
    ~& {\rm rank}({\bf W}_{\rm AB})=1, \label{com_rank_115}
\end{align}
\end{subequations}
where $\bar{k}_1=\frac{1}{\sigma^2_{\rm B}}$ and\\
${\bar{k}_2=\Big({\sum_{l=1}^L{\rm Tr}({\bf H}_{l{\rm E}}{\bf W}_{k}{\bf H}_{l{\rm E}}^{\rm H})+{\rm Tr}({\bf H}_{\rm CE}{\bf W}_{k}{\bf H}_{\rm CE}^{\rm H})+\sigma^2_{\rm E}}\Big)^{-1}}$. Equation \eqref{co_com_int_115} is a semi-definite programming (SDP) problem which was solved following conventional approach of neglecting the rank constraint in \eqref{com_rank_115}. Hence, by applying logarithm law, and rewriting the trace matrix with \cite[eq. 16]{matrix_cookbook}, the objective of \eqref{co_com_int_115} was written as fractional objective. Thereby enabling the use of Charnes-Cooper's transformation of the problem to \eqref{co_com_int_11b5}. Let $u=\big(1+{\rm Tr}({\bf H}_{\rm AE}^{\rm H}{\bf H}_{\rm AE}{\bf W}_{\rm AB})\big)^{-1}$, and ${\bf U}=u{\bf W}_{\rm AB}$, then \eqref{co_com_int_115} is equivalent to \eqref{co_com_int_11b5}.
\begin{subequations}\label{co_com_int_11b5}
\begin{align}
    \max_{{\bf U}, {\rm u}} ~&\big(u+{{\rm Tr}(\bar{k}_1{\bf H}_{\rm AB}^{\rm H}{\bf H}_{\rm AB}{\bf U})}\big),  \label{rs1b5} \\ 
    {\rm s.t.} ~& \big(u+{\bar{k}_1{\rm Tr}(\bar{k}_2{\bf H}_{\rm AE}^{\rm H}{\bf H}_{\rm AE}{\bf U})}\big)=1, \label{charnes_11b5} \\
    ~& {\rm Tr}(\bar{k}_2{\bf H}_{\rm AD}^{\rm H}{\bf H}_{\rm AD}u{\bf U})=0, \label{null_com_11b5}\\
    ~& {\rm Tr}(u{\bf U})=u, \label{com_power_11b5}
\end{align}
\end{subequations}
Equation \eqref{co_com_int_11b5} is convex and is easily solved with CVX \cite{cvx1}. We note that the rank constraint is dropped in \eqref{co_com_int_11b5} to allow for SDP solution. However, when the solution is obtained, the rank constraint was enforced using rank reduction technique like randomization if ${\rm rank}({\bf W}_{\rm AB})\neq 1$.

Furthermore, the sub-problem in terms of ${\bf W}_k$ was presented in \eqref{co_com_int_125}. This problem solves for the optimal weights of the radar transmitter to increase the average secrecy capacity of the setup demonstrated in fig.~\ref{case_25}.
\begin{subequations}\label{co_com_int_125}
\begin{align}
    \max_{{\bf W}_{k}} ~& \log_2\big(1+\gamma_{\rm B}\big)
    -\log_2\big(1+\hat{\gamma}_{\rm E}\big), \label{rs25} \\
    {\rm s.t.} ~&\log_2\Bigg(1+\frac{\sum_{l=1}^L{\rm Tr}({\bf H}_{l{\rm D}}{\bf W}_{k}{\bf H}_{l{\rm D}}^{\rm H})}{\sigma^2_{\rm D}}\Bigg)\geq r_{\rm th}, \label{radar_rate_threshold_125} \\
    ~& \sum_{l=1}^L{\rm Tr}({\bf H}_{l{\rm B}}{\bf W}_{k}{\bf H}_{l{\rm B}}^{\rm H})+{\rm Tr}({\bf H}_{{\rm CB}}{\bf W}_{k}{\bf H}_{{\rm CB}}^{\rm H})=0, \label{com_int_canc_125} \\
    ~& {\rm Tr}({\bf W}_{k})=1, \label{rad_power_125}\\
    ~& {\rm rank}({\bf W}_{k})=1. \label{rad_rank_125}
\end{align}
\end{subequations}
If we ignore the constant terms in the objective function that do not influence the optimization, with some mathematical manipulations, we obtain an SDP problem. The rank constraint was resolved using the technique described above. Hence, a convex equivalent of \eqref{co_com_int_125} was obtained and shown in \eqref{co_com_int_12b5}. Equation \eqref{co_com_int_12b5} is convex and can be solved with CVX \cite{cvx1}.
\begin{subequations}\label{co_com_int_12b5}
\begin{align}
    \max_{{\bf W}_{k}} ~&\log_2\Bigg(1-\frac{y3}{y1+y2+y3+\sigma^2_{\rm E}}\Bigg), \label{rs2b5} \\
    {\rm s.t.} ~&\log_2\Bigg(1+\frac{\sum_{l=1}^L{\rm Tr}({\bf H}_{l{\rm D}}{\bf W}_{k}{\bf H}_{l{\rm D}}^{\rm H})}{\sigma^2_{\rm D}}\Bigg)\geq r_{\rm th}, \label{radar_rate_threshold_12b5} \\
    ~& \sum_{l=1}^L{\rm Tr}({\bf H}_{l{\rm B}}{\bf W}_{k}{\bf H}_{l{\rm B}}^{\rm H})+{\rm Tr}({\bf H}_{{\rm CB}}{\bf W}_{k}{\bf H}_{{\rm CB}}^{\rm H})=0, \label{com_int_canc_12b5} \\
    ~& {\rm Tr}({\bf W}_{k})=1 \label{rad_power_12b5}.
\end{align}
\end{subequations}
where ${y1=\sum_{l=1}^L{\rm Tr}({\bf H}_{l{\rm E}}{\bf W}_{k}{\bf H}_{l{\rm E}}^{\rm H})}$, ${y2={\rm Tr}({\bf H}_{\rm CE}{\bf W}_{k}{\bf H}_{\rm CE}^{\rm H})}$, $y3={\rm Tr}({\bf H}_{\rm AE}{\bf W}_{{\rm AB}}{\bf H}_{\rm AE}^{\rm H})$.

\begin{figure}[H]
\centering
\includegraphics[width=1.0\linewidth]{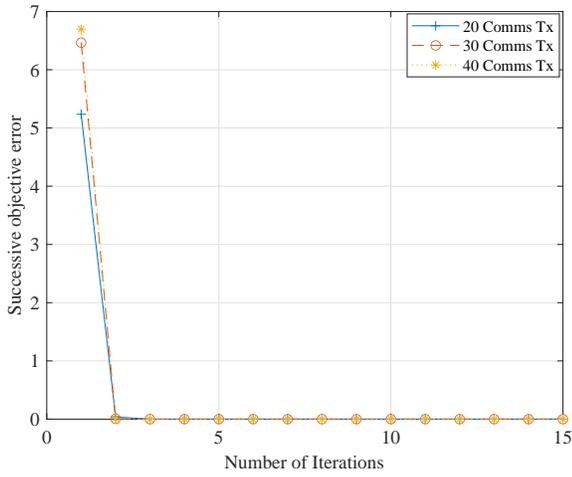}
\caption{Convergence of the sensing Algorithm~\ref{algo15}.}
\label{fig_conv5}
\end{figure}

In summary, the procedure to solve \eqref{co_com_int_1b5} follows Algorithm~\ref{algo15}. We note that $m_{max}$ is the maximum number of iterations. If the iterations terminates at the maximum number, then convergence was not obtained and the solution to the problem fails. However, it was shown in fig.~\ref{fig_conv5} that the Algorithm~\ref{algo15} always converge within a few number of simulations.
\begin{algorithm} [H]
    \caption{SCA Iterative algorithm for solving ${\bf W}_{\rm AB}$, ${\bf W}_{\rm AB}^0$ and ${\bf W}_k$}
  \begin{algorithmic}[1] \label{algo15}
    \STATE Initialize ${\bf W}_{{\rm AB}}^0$, ${\bf W}_k^0$ and $R_s^{0}$ such that the constraints in \eqref{co_com_int_1b5} were satisfied.
    \STATE $m \leftarrow 1.$
    \STATE \textbf{repeat}
    \begin{ALC@g}
    \STATE Compute and update ${\bf W}_{\rm AB}^m$ with \eqref{co_com_int_11b5}.
    \STATE Using updated ${\bf W}_{\rm AB}^m$, update ${\bf W}_k^m$ with \eqref{co_com_int_12b5}.
    \STATE Compute $R_s^{m}$ as defined in \eqref{rs_main5}.
    \STATE $\epsilon=\bigg|\frac{R_s^{m}-R_s^{m-1}}{R_s^{m}}\bigg |$.
    \STATE $m \leftarrow m + 1.$
    \end{ALC@g}
    \STATE \textbf{until} {$\epsilon \leq 10^{-5}$ OR $m\geq m_{max}$.}
    \STATE \textbf{Output:} ${\bf W}_{\rm AB}={\bf W}_{\rm AB}^m$ and ${\bf W}_k={\bf W}_k^m$.
  \end{algorithmic}
\end{algorithm}

\subsection{Uncooperative Systems}

\begin{figure}[H]
\centering
\includegraphics[width=1.0\linewidth]{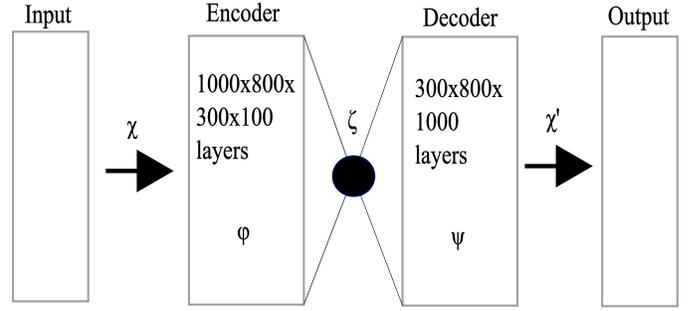}
\caption{Layer interaction of the autoencoder}
\label{fig_autoencoder5}
\end{figure}

When the radar and communications systems are uncooperative, the channel impulse responses are unknown. Hence relying on bemforming weights as an interference mitigation approach is insufficient. This is because nulling the channels as carried out in \eqref{null_com5} and \eqref{null_rad5} cannot be performed without knowledge of the channel impulse response. Therefore, to mitigate the cross interference of such radar communication systems, we implement a filter technology using autoencoder as shown in fig.~\ref{fig_autoencoder5}. Autoencoders use feature extraction to learn the variability of a multi-dimensional noisy data. The extraction is used to determine the noiseless version of the input data \cite{Papageorgiou_2020}. The noisy data referred to in this work include the desired signals, cross interference signals and AGWN. During the training, the network is configured to minimize the reconstruction loss given as
\begin{equation*}
    \mathcal{L}(\chi,\chi^{'})=||\chi-\chi^{'}||^2,
\end{equation*}
where $\chi^{'}=\psi({\bf W}^{'}\phi({\bf W}\chi+{\bf b})+{\bf b}^{'})$, $\{W,b\}$, $\{W^{'},b^{'}\}$ are the pair weights and biases for the encoder and decoder parts of the autoencoder. $\phi$ and $\psi$ are the encoder and decoder activation function respectively.

The neural network is first trained and validated with contextualized pilot data to extract features of the channel response and noise impact. Each snapshot of data comprise of 20,000 variations of a communication pilot signal and the radar target reflections. The layers of the autoencoders are activated using rectifier units. During the training phase, $10\%$ of the training data was used for validation. When the training phase is complete, the JCR signals are processed with the trained autoencoder networks at the communication and radar receivers. Since this neural network is domicile at the receivers, increase number of communication users or radar target does not affect its functionality.

\section{Results and Discussions} \label{results5}
\begin{figure}[H]
\centering
\includegraphics[width=1.0\linewidth]{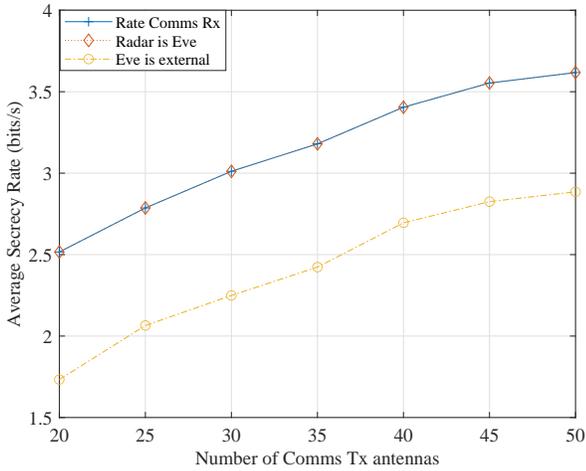}
\caption{PLS comparison with the optimal solution to \eqref{co_com_int5}.}
\label{fig_opt5}
\end{figure}

\begin{figure}[H]
\centering
\includegraphics[width=1.0\linewidth]{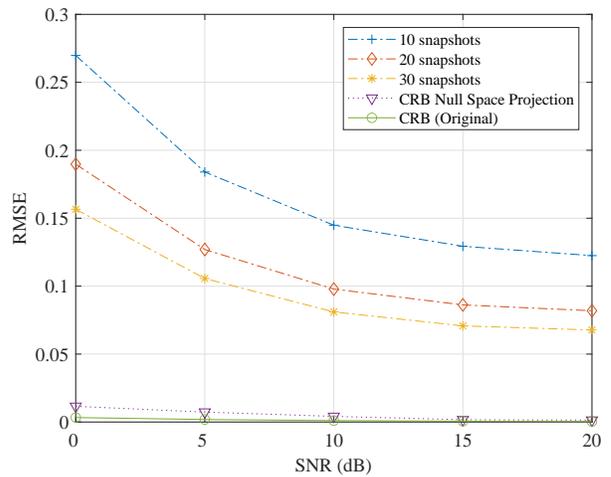}
\caption{RMSE performance graph of test Communication signals.}
\label{fig_rmse_comms5}
\end{figure}

\begin{figure}[H]
\centering
\includegraphics[width=1.0\linewidth]{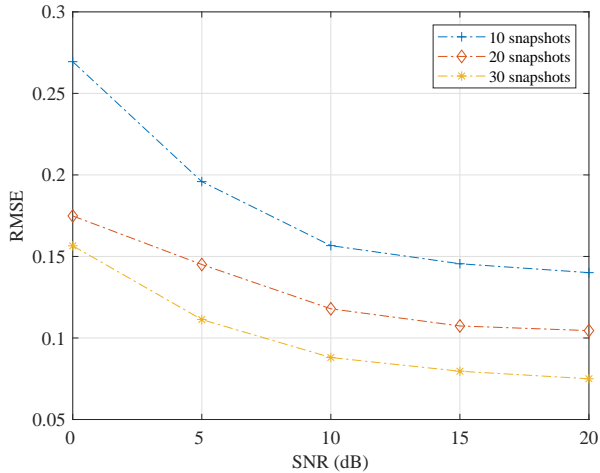}
\caption{RMSE performance graph of test radar reflected signals.}
\label{fig_rmse_RADAR5}
\end{figure}

The performance evaluation of the techniques discussed herein were obtained via MatLab simulations. The legends in fig.~\ref{fig_opt5} describes the number of radar antennas and the rate under consideration. In  figs.~\ref{fig_rmse_comms5} and \ref{fig_rmse_RADAR5}, the legends describes the number of snapshots used in training the network, whereas 'CRB Null Space Projection' and 'CRB (Original)' presents the null space projection given in \cite{Sodagari_2012} and the Cramer Rao lower bound respectively given known impulse response. The values of other simulation parameters are given in table \ref{table_par5}.

\begin{table}[!ht]
\begin{center}
\caption{Parameter description of the JCR model}\label{table_par5}
    \begin{tabular}{ | l | l | l |}
   
    \hline
    \textbf{Simulation parameter} & \textbf{Symbol} & \textbf{Value} \\ \hline
    Number of communication transmit antennas & $N_{\rm A}$ & $30$ \\ \hline
    Number of communication receive antennas & $N_{\rm B}$ & $4$ \\ \hline
    Number of radar transmit antennas & $N_{\rm C}$ & $30$ \\ \hline
    Number of radar receive antennas & $N_{\rm D}$ & $4$\\ \hline
    Carrier frequency (Surveillance) & $f_c$ & $2$GHz \\ \hline
    Distance between antenna elements & $d_x$ & $\frac{\lambda}{2}$\\ \hline
    Number of reflecting targets & $L$ & $3$ \\ \hline
    Noise power & $\sigma_{\rm B}^2$ and $\sigma_{\rm D}^2$ & $30$dBm \\ \hline
    Number of channel realizations &  & $500$\\ \hline
    \end{tabular}
\end{center}
\end{table}

In fig.~\ref{fig_opt5}, the cooperative system analysis of the communication and radar systems developed in section \ref{cooperative5} was presented. Recall that for a cooperative system, the channel information of the radar and communication systems are shared, and thereby effectively considered in the beamforming designs. From \eqref{null_com5} and \eqref{null_rad5}, the cross interference from both communication and radar systems are suppressed with the choice of the beamforming weights. The effectiveness of the beamforming designs allude the observation in fig.~\ref{fig_opt5} that although the radar receivers are illegitimate listeners to the communication signal, the average secrecy rate was shown to be approximately the same with the rate of the legitimate communication receiver. The rate of the legitimate communication receiver defines the upper bound to the average secrecy rate. In addition, we note that when the eavesdropper is not the radar receiver, thereby existing in an uncertain region, the average secrecy rate becomes smaller. This is because the beamforming designs does not null the external eavesdropper channel since it is unknown. However, even in the external eavesdropper scenario, the secrecy rate is positive and increasing with increasing number of transmission antennas. This result is expected since the cohabiting signal continue to offer interference at receivers not classified as legitimate. Additionally, from fig.~\ref{fig_opt5}, it was observed that while increase in the number of communication transmit antennas improves the performance of the systems in terms of transmission rate and average secrecy rate, the number of radar antennas has little/no impact. This is because the interference caused by the radar transmission was effectively removed by proper design of beamforming weights. 
\begin{figure}[H]
\centering
\includegraphics[width=1.0\linewidth]{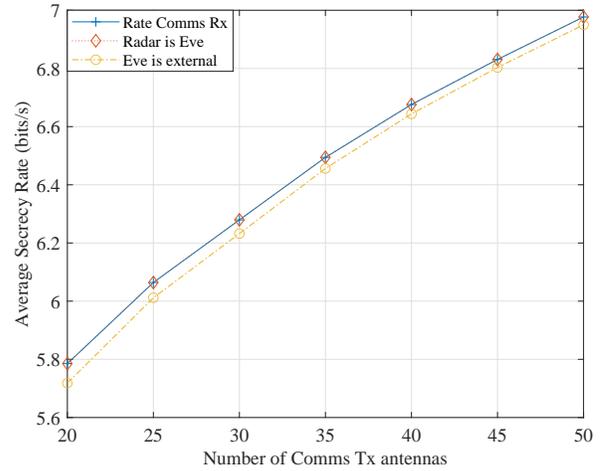}
\caption{PLS comparison with the optimal solution to \eqref{co_com_int_15}.}
\label{fig_av_sec_rate5}
\end{figure}

Furthermore we consider the PLS analysis of the communication/radar cohabiting system when PLS is considered in the design of the transmitter and receiver parameters as shown in fig.~\ref{fig_av_sec_rate5}. It is clear from fig.~\ref{fig_av_sec_rate5} that when the eavesdropper is not the radar receiving antenna, high PLS was obtained. Such high PLS was closer to the maximum rate when compared to the observations on fig.~\ref{fig_opt5}. This further supports the claim that by exploring techniques to cancel the interfering signal arising from cohabitation leads to higher PLS.

Figures~\ref{fig_rmse_comms5} and \ref{fig_rmse_RADAR5} illustrates the root mean square error (RMSE) of the interference cancelled received signal of the communication and radar signals respectively. For both figures, higher signal-to-noise ratio (SNR) presents lower decoding error (in terms of RMSE). By increasing the number of snapshots taken to train the neural network, better performance is also observed in both cases respectively. Specifically in fig.~\ref{fig_rmse_comms5}, the RMSE values were compared to the performance from the null projection from \cite{Sodagari_2012} and the Cramer Rao bound (CRB). We note that the channel response from \cite{Sodagari_2012} and CRB were known hence the lower RMSE. It is significant to observe that although better performance is observed with increasing snapshots, it is required that to approach the lower bounds, large snapshots are required. While this is desired, it places a constraint on the physical device used for training purposes. 

\section{Conclusion}\label{conclusion}
In conclusion, we assert that when full of part channel information are readily available, beamforming design mitigates the interference of radar and communication signals. However, under practical scenarios, where the channel information are usually indeterminate, we investigated a neural network based approach of filtering interfering and noise signals from the JCR system. We showed that sizeable chunk of training data guarantees better interference and noise cancellation when test data were deployed. This invariably leads to efficient cohabitation of the communication and radar systems. Therefore, the interference and noise cancellation approach is applicable when it is infeasible to estimate the channel properties of the JCR system as obtainable in real-time applications.

\bibliographystyle{IEEEtran}
\footnotesize{

\bibliography{IEEEabrv, ref_AIRS}}%
\end{document}